**Single-antenna super-resolution positioning with nonseparable toroidal pulses**

Ren Wang[1,2*], Pan-Yi Bao[1], Bing-Zhong Wang[1] and Yijie Shen[3,4*]

[1] *Institute of Applied Physics, University of Electronic Science and Technology of China, Chengdu 611731, China*

[2] *Yangtze Delta Region Institute (Huzhou), University of Electronic Science and Technology of China, Huzhou 313098, China*

[3] *Division of Physics and Applied Physics, School of Physical and Mathematical Sciences, Nanyang Technological University, Singapore 637378, Singapore*

[4] *Centre for Disruptive Photonic Technologies, The Photonics Institute, Nanyang Technological University, Singapore 637378, Singapore*

\* E-mail: rwang@uestc.edu.cn (R.W); yijie.shen@ntu.edu.sg (Y.S.)

## Abstract

**The fundamental principle of satellite/node-based positioning involves triangulating the receiver's coordinates through the intersection of spatial distances. Recent advancements in hybrid wireless networks have yielded high-precision positioning at decimetre-level (wavelength-level) [*Nature* 611, 473–478 (2022)], approaching the resolution limits in free space. Here, we present a three-dimensional (3D) super-resolution positioning paradigm in free space by utilizing a novel kind of topologically structured pulses, toroidal electromagnetic pulses [*Nat. Photonics* 16(7), 523-528 (2022); *Sci. Adv.* 10(2), eadl1803 (2024)]. Excited by the recent compact generator of toroidal pulses and their sophisticated topological and nonseparable structures, we demonstrate that the space-time nonseparability and skyrmion topology inherent in toroidal pulses can be harnessed to achieve freespace microwave 3D positioning with super-resolution accuracy, reaching the centimeter level, using a single emitting antenna. This work opens up new avenues for exploring the potential applications of topological electromagnetic pulses including but not limited to positioning, imaging, and sensing technologies.**

## Introduction

Freespace wireless microwave positioning plays a crucial role in daily life and scientific research [1-4]. Satellite navigation systems, such as the Global Positioning System (GPS), have revolutionized global navigation, mobile communication, and automated driving [5-7]. Recent advancements have led to the development of node network-based positioning systems by replacing satellite transmitters with ground-based stations or network nodes, thereby enhancing the precision and adaptability of positioning systems [8-11]. Both node network-based and satellite-based positioning systems rely on the multi-transmitter positioning paradigm and spatial distance intersection principle, illustrated in Fig. 1(a). However, two challenges, accuracy and node number, pose significant hurdles in freespace microwave positioning scenarios. Firstly, the precision of positioning based on the spatial distance intersection principle depends on synchronization and ranging. Typically, the synchronization accuracy of GPS systems reaches several nanoseconds, resulting in meter-level positioning errors. Recent advancements in hybrid wireless networks have pushed synchronization precision to subnanosecond levels, enabling decimeter-level (wavelength-level) high-accuracy positioning [10,11]. This development approaches the resolution limits in free space and the accuracy is difficult to further enhance. Secondly, when accurately measuring flying time or distance $l$, each transmitter can only ascertain that the receiver lies on a sphere with a radius of $l$. Consequently, it is impossible to determine the 3D coordinates of a target using a single transmitting antenna based on the spatial distance intersection principle.

Topological electromagnetic fields exhibit vast potential applications in information and energy transmission [12-18] and super-resolution metrology or microscopy [19-24], thereby offering a potential avenue for achieving precise 3D positioning based on a single transmission antenna in free space. In 1996, Hellwarth and Nouchi introduced toroidal pulses, an exact solution to Maxwell's equations [25], which possess complex non-transverse skyrmion topological structures [26-29], space-time nonseparability [30-33], and strong interactions with matter [34,35]. Recently, electromagnetic toroidal pulses have been successfully generated in the optics, terahertz, and microwave [36-38], laying the foundation for their practical applications. However, there have been no reports of successful experimental implementations of toroidal pulses until now.

In this paper, we present a novel method for freespace microwave 3D positioning based on the space-time nonseparability and skyrmion topology of toroidal pulses, enabling 3D spatial positioning with super-resolution accuracy, reaching the centimeter level, using a single emitting antenna. The effectiveness of this method is validated through both theoretical analysis and experimental verification. As a result, the proposed methodology utilizing toroidal pulses lays the groundwork for a new positioning paradigm.

**Results**

**Positioning principle based on nonseparable toroidal pulses.** The space-time nonseparability and space-polarization nonseparability of toroidal pulses, evident in their analytical expressions, are the fundamental principles underlying the proposed 3D positioning method. The expression for the field distribution of transverse-magnetic (TM) toroidal pulses (Figs. 1(b1 and b2)) in a cylindrical coordinate system is as follows [25]:

$$H_\theta = 4if_0 \frac{\rho(q_1+q_2-2ict)}{[\rho^2+(q_1+i\tau)(q_2-i\sigma)]^3} \quad (1)$$

$$E_\rho = 4if_0\sqrt{\frac{\mu_0}{\varepsilon_0}} \frac{\rho(-q_1+q_2-2iz)}{[\rho^2+(q_1+i\tau)(q_2-i\sigma)]^3} \quad (2)$$

$$E_z = -4f_0\sqrt{\frac{\mu_0}{\varepsilon_0}} \frac{\rho^2-(q_1+i\tau)(q_2-i\sigma)}{[\rho^2+(q_1+i\tau)(q_2-i\sigma)]^3} \quad (3)$$

where, $H_\theta$ is the magnetic field component in the $\theta$ direction, $E_\rho$ and $E_z$ are the electric field components in the $\rho$ and z directions, respectively. The $i$ denotes the imaginary unit, $f_0$ represents a real constant, $q_1$ and $q_2$ are adjustable real positive parameters, $c$ represents the speed of light, $t$ denotes time, $\tau = z - ct$, $\sigma = z + ct$. $\mu_0$ and $\varepsilon_0$ represent the permeability and permittivity in vacuum, respectively.

Due to the inability to separate the spatial variable $\rho$ from the temporal variable $t$ in the expression of toroidal pulses mentioned above, their field equations cannot be expressed as spatial field forms with the temporal variable eliminated or temporal field forms with the spatial variable eliminated. This unique characteristic is referred to as space-time nonseparability, which is equivalent to space-frequency nonseparability. Due to their space-time nonseparability, toroidal pulses exhibit distinct electromagnetic fields in both the temporal and frequency domains at each spatial location [32]. When examining the maximum field distribution of toroidal pulses at a specific frequency point as it varies with the propagation space, we obtain a distribution of maximum spectral lines as shown in Fig. 1(b3). The solid lines represent the radial maximum values of the $E_\rho$ component along the z-axis for several frequencies ranging from $f_{max}$ to $f_{min}$, with decreasing frequency from the inner to the outer regions. The maximum spectral lines are mutually separated and never intersect, which is one manifestation of the space-time nonseparability of toroidal pulses [32]. This spatial-frequency correspondence forms the foundation for positioning based on the space-time nonseparability. The correspondence between spatial and frequency spectra enables the determination of target coordinates ($\rho$, z). For further details, please refer to the supplementary material.

Drawing inspiration from animals that navigate using electromagnetic wave polarization [39,40], we can also utilize the polarization properties of toroidal fields for positioning purposes. Toroidal fields exhibit space-polarization nonseparability apart from space-time nonseparability [26-29], with their skyrmion polarization textures having been experimentally observed [38]. This intricate spatial topological feature holds the potential for achieving ultrahigh resolution positioning. Taking TM toroidal pulses as an example, $E_\rho$ exhibits skyrmion-type polarization, as indicated in Fig. 1(b4). The azimuthal coordinate $\theta$ of the target can be determined based on the polarization characteristics of the electric field vector. Consequently, the time/frequency-polarization information at each 3D spatial point within the primary distribution region of the toroidal field is unique, meaning that it can be uniquely characterized by the time-polarization state or frequency-polarization state. As shown in Fig. 1, by discretizing the 3D space into a grid and placing receivers at each grid point, we can measure the time/frequency-polarization state at each point and store it along with the corresponding spatial coordinates as a positioning information database. Subsequently, by measuring the time/frequency-polarization state at any target spatial point and comparing it with the positioning information database, we can obtain the corresponding spatial coordinates, thus achieving target positioning.

**Single-antenna 3D positioning performances.** Shen et al. introduced the space and time states as a set of entangled states based on the spatial and temporal properties of toroidal pulses [32]. These states were quantized through measurements of the state density matrix to assess the spatiotemporal inseparability of the toroidal pulses. Inspired by this approach, we define the frequency-polarization amplitude state $|\psi_a\rangle$ and the phase state $|\psi_p\rangle$, leveraging the unique frequency-polarization information at each positioning within the propagation space of the toroidal pulses.

$$|\psi_a\rangle = [|E_{hor}(\lambda,\rho,z,\theta)|, |E_{ver}(\lambda,\rho,z,\theta)|] \tag{4}$$

$$|\psi_p\rangle = [\arctan\left(\frac{Im(E_{hor}(\lambda,\rho,z,\theta))}{Re(E_{hor}(\lambda,\rho,z,\theta))}\right), \arctan\left(\frac{Im(E_{ver}(\lambda,\rho,z,\theta))}{Re(E_{ver}(\lambda,\rho,z,\theta))}\right)] \tag{5}$$

where, $E_{hor}$ and $E_{ver}$ denote the horizontal and vertical components of the normalized transverse electric field of the toroidal pulses, respectively. We exploit the regularity of the spatial field distribution of the toroidal pulses through measurements of the amplitude of the electric field, and the regularity of its spatial polarization by measuring the phase angles of the electric field components in different directions. Specifically, the frequency-polarization amplitude state encodes information about spatial distance and radial position, while the frequency-polarization phase state contains information about the spatial azimuthal position. Drawing inspiration from the concept of fidelity in quantum mechanics, we introduce a spatial pseudo-spectrum tailored for receiver positioning.

$$F_{tol} = F_a * F_p \tag{6}$$

$$F_a = \frac{\langle\psi_{a,tar}|\psi_{a,pre}\rangle\langle\psi_{a,tar}|\psi_{a,pre}\rangle}{\langle\psi_{a,tar}|\psi_{a,tar}\rangle\langle\psi_{a,pre}|\psi_{a,pre}\rangle} \tag{7}$$

$$F_p = \frac{\langle\psi_{p,tar}|\psi_{p,pre}\rangle\langle\psi_{p,tar}|\psi_{p,pre}\rangle}{\langle\psi_{p,tar}|\psi_{p,tar}\rangle\langle\psi_{p,pre}|\psi_{p,pre}\rangle} \tag{8}$$

where, the subscript "tar" represents the frequency-polarization state measured by the receiver at the target location, while the subscript "pre" denotes the frequency-polarization state obtained from a pre-measured positioning information database at different positions within the area of interest. In Eqs. (6-8), both $F$ and $|\psi_{pre}\rangle$ are functions of the coordinate positioning ($\rho$, $z$, $\theta$), while $|\psi_{tar}\rangle$ is computed based on the signals received by the receiver at the location($\rho_0, z_0, \theta_0$). Consequently, the computation of $F_{tol}$ using the aforementioned equations yields pseudo-spectrum values for different coordinates ($\rho$, $z$, $\theta$), as illustrated in Fig. 2. In Fig. 2, the target is situated at ($x$, $y$, $z$) = (0, 0.1, 0.2) m, with the toroidal pulses propagating in the $z$-direction, and the frequency range used for computing the pseudo-spectrum is 2-10 GHz.

Fig. 2(a) depicts the theoretical pseudo-spectrum computed using ideal toroidal pulses (given by Eqs. (1-3) with $q_1$=0.02m and $q_2$=50$q_1$). It is evident from the figure that the spatial coordinate corresponding to the maximum pseudo-spectrum value, i.e. 1, aligns with the target location, with high pseudo-spectrum values observed near the target and low values at distant locations. This clearly indicates the target's coordinate. In practical positioning applications, however, test errors due to noise are inevitable. To assess the performance of our method in noisy conditions, we superimposed Gaussian white noise with a signal-to-noise ratio of 20 dB on the ideal toroidal pulses field at each position. The recomputed pseudo-spectrum is shown in Fig. 2(b), indicating a degree of noise robustness. To validate the positioning method proposed in this study, we conducted experimental tests using a coaxial horn toroidal pulses emitting antenna (Fig. 1(b1)), as described in [38]. Experimental settings are detailed in the Method. The measured 3D pseudo-spectrum is presented in Fig. 2(c), clearly identifing the target's coordinate in the experimental results, agreeing well with theoretical predictions.

As observed in Fig. 2, although the maximum value of $F_{tol}$ in the pseudo-spectrum indicates the target's position, the relatively high values of $F_{tol}$ in the vicinity of the target can potentially affect judgement. To improve the purity of the pseudo-spectrum and enhance resolution without compromising positioning accuracy, we can employ techniques to sharpen the pseudo-spectrum. These techniques are described in detail in the Supplementary Material. In addition, it is worth noting that due to the inherent regularity in the spatial-frequency-polarization entanglement of toroidal pulses, we can achieve accurate positioning using spatial pseudo-spectrum computations regardless of the type of receivers.

**Simultaneous positioning of multiple receivers.** Beyond the positioning of a single receiver, our method also enables the simultaneous positioning of multiple receivers situated at different locations. By separately computing pseudo-spectra using signals received by different receivers and superimposing the sharpened pseudo-spectra, we can obtain a pseudo-spectrum map for multi-target positioning. Since the pseudo-spectrum values outside the targets' positions are very small, the superimposed pseudo-spectra corresponding to individual receivers do not mutually obscure each other, as demonstrated in Fig. 3. In this figure, the coordinates of targets 1, 2, and 3 are $(r, z)$ = (0.1, 0.4) m, (0.2, 0.2) m, and (0.25, 0.65) m, respectively. Both numerical computations and experimental measurements yield pseudo-spectrum peaks that align with the coordinates of targets.

**Evaluation of positioning accuracy.** To quantitatively demonstrate the positioning performance of our method, we employed the displacement metric, which quantifies the spatial deviation between the estimated coordinate (derived from the peak of the spatial pseudo-spectrum) and actual target's coordinate. Fig. 4 illustrates the displacement when targets are situated at various locations within the primary coverage area of the toroidal pulses. Due to the rotational symmetry of toroidal pulses, examining half of the cylindrical coordinate system along the *z*-axis suffices to capture their full spatial characteristics. In Fig. 4, the value at each ($\rho_0$, $z_0$) coordinate represents the displacement normalized by the shortest wavelength $\lambda_{min} = 0.03m$ (corresponding to 10GHz) used in the pseudo-spectrum computation, when the actual target's coordinate is ($\rho_0$, $z_0$). When utilizing ideal toroidal pulses for positioning, the displacement is zero regardless of the target's coordinate. When noise is introduced to toroidal pulses, the displacement remains below $0.5\lambda_{min}$ (0.3 $\lambda_{cen}$ at the central frequency 6 GHz) for 91.82% of the coordinates, indicating robust positioning accuracy even in noisy environments. Experimental results show that 97.15% of the cases are less than $0.5\lambda_{min}$ (0.3 $\lambda_{cen}$) when targets are situated at different locations, consistent with the findings obtained using noisy toroidal pulses. In conclusion, owing to space-time nonseparability and skyrmion topology, positioning methods based on toroidal pulses can achieve super-resolution accuracy.

**Discussion**

By harnessing the inherent space-time nonseparability and skyrmion topology of toroidal pulses, we have realized super-resolution 3D positioning of receivers through the utilization of a solitary toroidal pulse emitting antenna in free space, making the method applicable for various applications including global navigation, mobile communication, automated driving, and others.

It's worth noting that scholars have proposed numerous super-resolution imaging or positioning methods [41-43], but their application scenarios differ from freespace microwave positioning. For example, by controlling multipath scattering in reverberation chambers, it is conceivable to achieve super-resolution localization [44-46]. However, this method is unsuitable for freespace microwave positioning, where reverberation scattering is unavailable for exploitation.

In addition to the cooperative transmitter-receiver positioning results shown in this paper, the nonseparability and skyrmion topology of toroidal pulses holds promise for pioneering applications in radar detection, internet of things, and penetrative imaging. For scattering scenarios, employing a single coaxial horn antenna to emit toroidal pulses and employing broadband dual-polarization antennas to receive scattered target echoes allows for position determination based on echo polarization and spectral characteristics. By integrating neural network techniques with toroidal pulses, it may even be possible to reconstruct high-precision 3D shapes of targets. Radar or imaging systems based on toroidal pulses offer the potential for higher resolution with simpler system architectures and fewer transceiver units.

As a higher-order form of toroidal pulses, supertoroidal pulses exhibit more pronounced variations in spatial frequency distribution and polarization distribution [27]. Therefore, utilizing supertoroidal pulses holds promise for constructing higher-resolution positioning, detection, imaging, and other systems. Moreover, supertoroidal pulses may possess non-diffracting propagation characteristics [28], potentially enabling systems based on them to achieve greater effective detection ranges.

As classical electromagnetic waves, toroidal pulses exhibit identical propagation and topological characteristics across microwave, terahertz, and optical frequency bands.

Although this paper primarily discusses the application of toroidal pulses in microwave positioning systems, toroidal pulses have also been generated in the optical and terahertz frequencies [36,37]. Hence, the methodology proposed in this paper also holds potential for applications in optical fields such as nanoscale particle metrology and super-resolution imaging.

**Methods**

**Numerical positioning method for ideal toroidal pulses.** The specific implementation steps of numerical positioning based on ideal toroidal pulses are as follows: (1) Generate an ideal toroidal pulse based on Eqs. (1-3) in Matlab software with $q_1$=0.02m and $q_2$=50$q_1$, setting the propagation axis of the toroidal pulse as the z-axis; (2) Uniformly sample within the region of $x$=[-0.4 0.4]m, $y$=[-0.4 0.4]m, $z$=[0 0.8]m with a 1 cm interval, and record the spectrum of the toroidal pulse at each sampling point; (3) Assume the target is located at Position 0 ($\rho_0, z_0, \theta_0$), use the data recorded at Position 0 ($\rho_0, z_0, \theta_0$) and Position 1 ($\rho_1, z_1, \theta_1$) obtained in step (2) to calculate the pseudo-spectrum value $F_{tol}(\rho_1, z_1, \theta_1)$ at Position 1 ($\rho_1, z_1, \theta_1$) according to Eqs. (4-8); (4) Traverse the coordinates of Position 1 throughout the entire computational domain, repeatedly perform Step (3), and obtain the pseudo-spectrum map of the entire space; (5) Identify the position ($\rho_0', z_0', \theta_0'$) with the highest amplitude in the pseudo-spectrum map, which corresponds to the determined target position; (6) Calculate the distance between the determined position ($\rho_0', z_0', \theta_0'$) and the actual target position ($\rho_0, z_0, \theta_0$), which represents the positioning displacement value for the target position ($\rho_0, z_0, \theta_0$); (7) Traverse the target Position 0 throughout the entire space, repeatedly perform Steps (3-6), and obtain the positioning displacement value corresponding to each position in the entire space.

**Numerical positioning method for noisy toroidal pulses.** The specific implementation steps of numerical positioning in the presence of noise are as follows: (1) Follow the same procedure as in the case for ideal toroidal pulse; (2) After step (2) in the ideal toroidal pulse scenario, add two sets of random Gaussian white noise with signal-to-noise ratios of 20dB to the signals at each position, respectively referred to as pre-measured and target data sets; (3) Assume the target is located at Position 0 $(\rho_0, z_0, \theta_0)$, use the data from Position 0 $(\rho_0, z_0, \theta_0)$ in the target data set recorded in step (2) and the data from Position 1 $(\rho_1, z_1, \theta_1)$ in the pre-measured data set according to Eqs. (4-8) to calculate the pseudo-spectrum value $F_{tol}(\rho_1, z_1, \theta_1)$ at Position 1 $(\rho_1, z_1, \theta_1)$; (4) Traverse the coordinates of Position 1 in the pre-measured data set throughout the entire computational domain, repeatedly perform Step (3), and obtain the pseudo-spectrum map of the entire space; (5-6) Follow the same steps as in the ideal toroidal pulse scenario; (7) Traverse the target Position 0 throughout the entire space in the target data set, repeatedly perform Steps (3-6), and obtain the positioning displacement value corresponding to each position in the entire space.

**Experimental positioning method.** The TM toroidal pulses transmitter employed in the experimental validation of the proposed positioning technology is a coaxial horn antenna proposed in Ref. [38]. The experimental setup utilized coaxial horn antennas operating within the frequency range of 1.3-10 GHz as the transmitting antennas for toroidal fields, while double-ridged waveguide horn antennas (operating within 1-18 GHz) served as the dual-polarized receiving antennas, simulating the user to be localized. The geometric center of the coaxial horn antenna aperture was designated as the origin of the cylindrical coordinate system. The experimental frequency range was set to 2-10 GHz, with the spatial *z*-axis originated at a distance of 0.2 m from the coaxial horn antenna. Under identical experimental conditions, the spatial $E_\rho$ distribution was measured twice. The amplitude and phase deviations between the two experimental datasets were found to be within 1%-2%. The experimental system was situated in a planar field measurement chamber, as depicted in Fig. 5(a), where the transmitting and receiving antennas were connected to a vector network analyzer and controlled by a computer. The double-ridged waveguide horn antenna could be moved in 3D space using a rail system, simulating various user coordinates. Fig. 5(b) illustrates the cross-sectional structure of the coaxial horn antenna, where a 180° rotation along the symmetry axis reveals the complete antenna geometry. The black color represents the metal conductor, while the green color indicates a material with a dielectric constant of 1.3, serving to support the inner and outer conductors of the coaxial horn antenna. The antenna is fed through a coaxial cable connected at its base.

The procedure for obtaining experimental positioning results is as follows: (1) A 2-10 GHz signal is fed into the coaxial horn toroidal field transmitter, with the central axis of the coaxial horn set as the z-axis and the aperture of the horn set at *z*=0; (2) Spectra are measured at each spatial position within the region of *x*=[-0.4 0.4]m, *y*=[-0.4 0.4]m, *z*=[0 0.8]m at 1 cm intervals, and recorded as the pre-measured data set; (3) The target under test is placed at Position 0 $(\rho_0, z_0, \theta_0)$, and the spectrum is re-measured, representing the received signal spectrum when the target is located at Position 0. Using this re-measured spetrum and the data from Position 1 $(\rho_1, z_1, \theta_1)$ in the pre-measured data set, the pseudo-spectrum value $F_{tol}(\rho_1, z_1, \theta_1)$ at Position 1 $(\rho_1, z_1, \theta_1)$ is calculated according to Eqs. (4-

8); (4-6) Follow the same steps as in the case of positioning with noisy toroidal pulse; (7) The target under test is placed at different positions through the entire space, repeating Steps (3-6) continuously to obtain the positioning displacement value corresponding to each position.

**References**

1. Prescott, W. H., Davis, J. L., & Svarc, J. L.. Global positioning system measurements for crustal deformation: Precision and accuracy. *Science* **244**(4910), 1337-1340 (1989).
2. Giles, J.. EU plans global positioning system. *Nature* **410**(6831), 853-853 (2001).
3. Elsheikh, M., Iqbal, U., Noureldin, A., & Korenberg, M.. The implementation of precise point positioning (PPP): a comprehensive review. *Sensors* **23**(21), 8874 (2023).
4. Kentosh, J. , & Mohageg, M. . Global positioning system test of the local position invariance of planck's constant. *Phys. Rev. Lett.* **108**(11), 110801 (2012).
5. Rizos, C., & Yang, L.. Background and recent advances in the Locata terrestrial positioning and timing technology. *Sensors* **19**(8), 1821 (2019).
6. Humphreys, T. E., Murrian, M. J., & Narula, L.. Deep-urban unaided precise global navigation satellite system vehicle positioning. *IEEE Intell. Transp. Syst. Mag.* **12**(3), 109-122 (2020).
7. Giovannetti, V. , Lloyd, S. , & Maccone, L.. Quantum-enhanced positioning and clock synchronization. *Nature* **412**(6845), 417-419 (2001).
8. del Peral-Rosado, J. A., Raulefs, R., López-Salcedo, J. A., & Seco-Granados, G.. Survey of cellular mobile radio localization methods: From 1G to 5G. *IEEE Commun. Surveys Tuts.* **20**(2), 1124-1148 (2017).

9. Janssen, T., Koppert, A., Berkvens, R., & Weyn, M.. A survey on IoT positioning leveraging LPWAN, GNSS and LEO-PNT. *IEEE Internet Things J.* **10**(13), 11135-11159, (2023).

10. Hui Chen, et al. Phone signals can help you find your way in cities even without GPS. *Nature* **611**, 454-455 (2022).

11. Koelemeij, J.C.J., Dun, H., Diouf, C.E.V. et al. A hybrid optical–wireless network for decimetre-level terrestrial positioning. *Nature* **611**, 473–478 (2022).

12. Nye, J. F. & Berry, M. V. Dislocations in wave trains. In A Half-Century of Physical Asymptotics and Other Di-versions: Selected Works by Michael Berry, 6–31 (World Scientific, 1974).

13. Willner, A. E., Pang, K., Song, H., Zou, K., & Zhou, H.. Orbital angular momentum of light for communications. *Appl. Phy. Rev.* **8**(4), 041312, (2021).

14. Berry, M. Making waves in physics. *Nature* **403**, 21–21 (2000).

15. Bliokh, K. Y., Karimi, E., et al. Roadmap on structured waves. *J. Opt.* **25**(10), 103001 (2023).

16. Wan, Z., Wang, H., Liu, Q., Fu, X., & Shen, Y.. Ultra-degree-of-freedom structured light for ultracapacity information carriers. *ACS Photonics* **10**(7), 2149-2164 (2023).

17. Wan, Z., Shen, Y., Wang, Z. et al. Divergence-degenerate spatial multiplexing towards future ultrahigh capacity, low error-rate optical communications. *Light Sci Appl*. **11**, 144 (2022).

18. Pryamikov, A. Rising complexity of the OAM beam structure as a way to a higher data capacity. *Light Sci Appl*. **11**, 221 (2022).

19. Berg-Johansen, S., Töppel, F., Stiller, B., Banzer, P., Ornigotti, M., Giacobino, E., ... & Marquardt, C.. Classically entangled optical beams for high-speed kinematic sensing. *Optica* **2**(10), 864-868 (2015).

20. Yuan, G. H., & Zheludev, N. I. Detecting nanometric displacements with optical ruler metrology. *Science* **364**(6442), 771-775 (2019).

21. Vettenburg, T., Dalgarno, H. I., Nylk, J., Coll-Lladó, C., Ferrier, D. E., Čižmár, T., ... & Dholakia, K.. Light-sheet microscopy using an Airy beam. *Nat. Methods* **11**(5), 541-544 (2014).

22. Zheludev, N. I., & Yuan, G. Optical superoscillation technologies beyond the diffraction limit. *Nat. Rev. Phys.* **4**(1), 16-32 (2022).

23. Jia, S., Vaughan, J. C., & Zhuang, X.. Isotropic three-dimensional super-resolution imaging with a self-bending point spread function. *Nat. Photonics* **8**(4), 302-306 (2014).

24. Pei Miao et al. ,Orbital angular momentum microlaser.*Science* **353**, 464-467 (2016).

25. Hellwarth, R. W., & Nouchi, P.. Focused one-cycle electromagnetic pulses. *Phy. Rev. E* **54**(1), 889 (1996).

26. Shen, Y., Zhan. Q., et al. Roadmap on spatiotemporal light fields. *J. Opt.* **25**(9), 093001 (2023).

27. Shen, Y., Hou, Y., Papasimakis, N., & Zheludev, N. I. Supertoroidal light pulses as electromagnetic skyrmions propagating in free space. *Nat. Commun.* **12**(1), 5891 (2021).

28. Shen, Y., Papasimakis, N., & Zheludev, N. I.. Nondiffracting supertoroidal pulses: optical "Kármán vortex streets". *Nat. Commun.* (2024).

29. Kaelberer, T., Fedotov, V. A., Papasimakis, N., Tsai, D. P., & Zheludev, N. I.. Toroidal dipolar response in a metamaterial. *Science* **330**(6010), 1510-1512 (2010).

30. He, C., Shen, Y., & Forbes, A. Towards higher-dimensional structured light. *Light Sci Appl*. **11**(1) 205 (2022).

31. Shen, Y. & Rosales-Guzmán, C. Nonseparable states of light: from quantum to classical. *Laser Photon. Rev.* **16**(7), 2100533 (2022).

32. Shen, Y., Zdagkas, A., Papasimakis, N., & Zheludev, N. I. (2021). Measures of space-time nonseparability of electromagnetic pulses. *Phy. Rev. Res.* **3**(1), 013236.

33. Shen, Y., Zhang, Q., Shi, P., Du, L., Yuan, X., & Zayats, A. V.. Optical skyrmions and other topological quasiparticles of light. *Nat. Photonics* **18**(1), 15-25 (2024).

34. Papasimakis, N., Fedotov, V., Savinov, V. et al. Electromagnetic toroidal excitations in matter and free space. *Nat. Material* **15**, 263–271 (2016).

35. Raybould, T., Fedotov, V. A., Papasimakis, N., Youngs, I., & Zheludev, N. I.. Exciting dynamic anapoles with electromagnetic doughnut pulses. *Appl. Phy. Lett.* **111**(8), 081104 (2017).

36. Zdagkas, A., McDonnell, C., Deng, J., Shen, Y., Li, G., Ellenbogen, T., ... & Zheludev, N. I.. Observation of toroidal pulses of light. *Nat. Photonics* **16**(7), 523-528 (2022).

37. Jana, K., Mi, Y., Møller, S. H., Ko, D. H., Gholam-Mirzaei, S., Abdollahpour, D., ... & Corkum, P. B.. Quantum control of flying doughnut terahertz pulses. *Sci. Adv.* **10**(2), eadl1803 (2024).

38. Wang, R., Bao, P. Y., Hu, Z. Q., Shi, S., Wang, B. Z., Zheludev, N. I., & Shen, Y.. Free-space propagation and skyrmion topology of toroidal electromagnetic pulses. arXiv preprint arXiv:2311.01765 (2023).

39. Powell, S. B., Garnett, R., Marshall, J., Rizk, C., & Gruev, V.. Bioinspired polarization vision enables underwater geolocalization. *Sci. Adv.* **4**(4), eaao6841 (2018).

40. Graydon, O.. Global position by polarization. *Nat. Photonics* **12**(6), 318-318 (2018).

41. Power, R. M., Tschanz, A., Zimmermann, T., & Ries, J.. Build and operation of a custom 3D, multicolor, single-molecule localization microscope. *Nat. Protocols* **686** 1-59 (2024).

42. Astratov, V. N., Sahel, Y. B., Eldar, Y. C., Huang, L., Ozcan, A., Zheludev, N., ... & Lecler, S.. Roadmap on label‐free super‐resolution imaging. *Laser Photon. Rev.* **17**(12), 2200029 (2023).

43. Zheludev, N. I., & Yuan, G.. Optical superoscillation technologies beyond the diffraction limit. *Nat. Rev. Phys.* **4**(1), 16-32 (2022).

44. Cohen, S. D., Cavalcante, H. L. D. S., & Gauthier, D. J.. Subwavelength position sensing using nonlinear feedback and wave chaos. *Phys. Rev. Lett.* **107**(25), 254103 (2011).

45. Del Hougne, P. , Imani, M. F. , Fink, M. , Smith, D. R. , & Lerosey, G. . Precise localization of multiple noncooperative objects in a disordered cavity by wave front shaping. *Phys. Rev. Lett.* **121**(6), 063901 (2018).

46. Gigan, P. del Hougne, Deeply subwavelength localization with reverberation-coded aperture. *Phys. Rev. Lett.* **127**, 043903 (2021).

**Acknowledgements**

The authors acknowledge the supports of the the National Natural Science Foundation of China (62171081, 61901086, U2341207), the Aeronautical Science Foundation of China (2023Z062080002), and the Natural Science Foundation of Sichuan Province (2022NSFSC0039). Y. Shen also acknowledges the support from Nanyang Technological University Start Up Grant, Singapore Ministry of Education (MOE) AcRF Tier 1 grant (RG157/23), and MoE AcRF Tier 1 Thematic grant (RT11/23).

**Contributions**

R.W. conceived the ideas and supervised the project, R.W. and P.Y.B. performed the theoretical modeling and numerical simulations, R.W. developed the experimental methods, P.Y.B. and R.W. conducted the experimental measurements, R.W., P.Y.B. and Y.S. conducted data analysis. All authors wrote the manuscript and participated the discussions.

**Competing interests**

The authors declare no competing interests.

**Data and materials availability**

The data that support the findings of this study are available from the corresponding author upon reasonable request.

**Additional information**

**Supplementary information** is available for this paper. Correspondence and requests for materials should be addressed to R.W. and Y.S..

# Figures

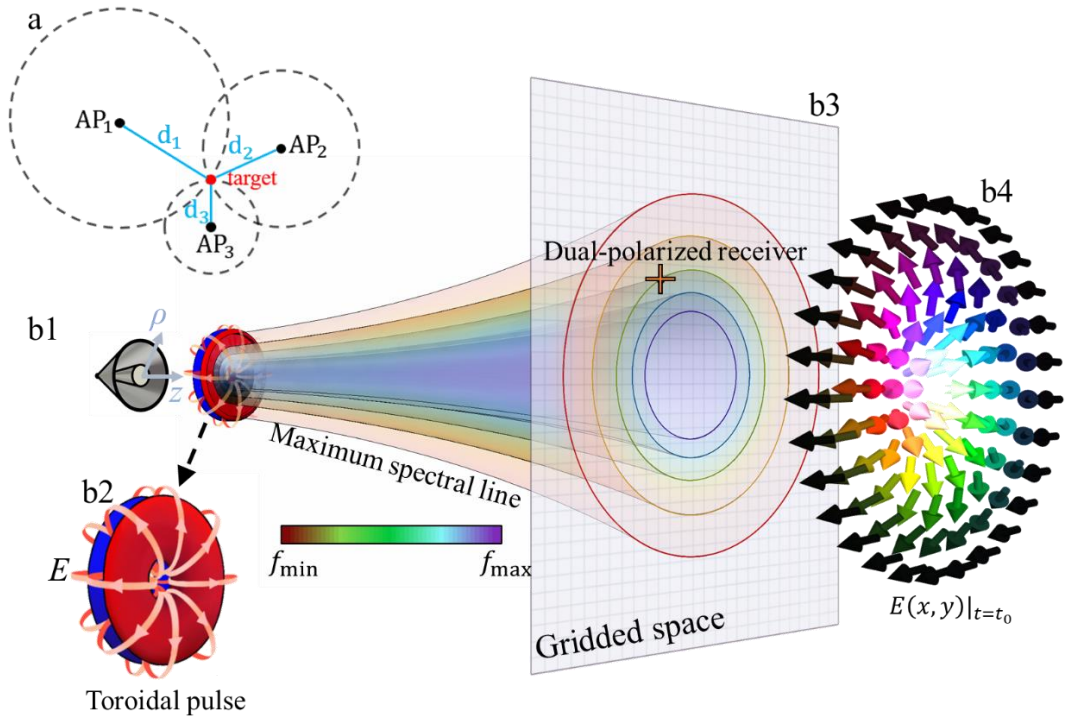

**Fig. 1 Schematic diagram of 3D positioning method based on toroidal pulses.** (**a**) Traditional spatial distance intersection principle, where AP represents access point; (**b1**) Toroidal pulses emission antenna; (**b2**) Electric field distribution of TM toroidal pulses, with the electric field $E$ indicated by encircling pink arrows; (**b3**) Schematic diagram of the maximum frequency positioning and polarization distribution of toroidal pulses. We annotate the maximum positioning of the single-frequency field distribution in the $\rho$ direction (referred to as the maximum spectrum line) using solid lines of different colors (from violet to red indicating from high to low frequencies). After spatial discretization, the maximum spectrum lines intersect with circular coils on a plane perpendicular to the propagation axis (z-axis). (b4) Skyrmion topology in toroidal pulses.

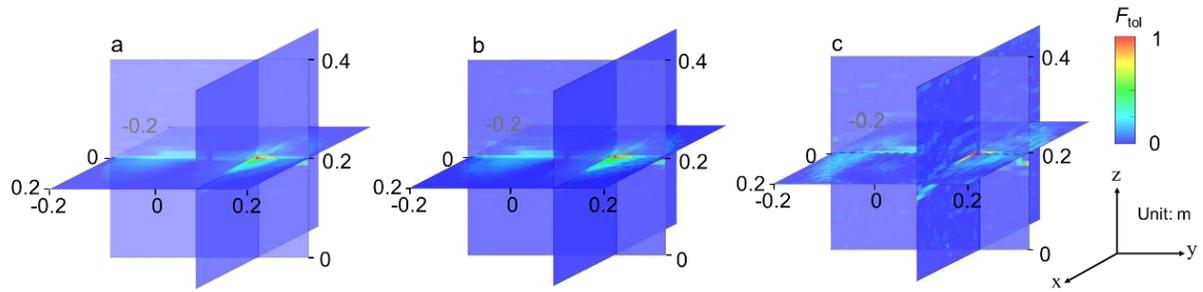

**Fig. 2 Three-dimensional positioning pseudo-spectrum based on toroidal pulses.** (a) Ideal toroidal pulse-based positioning pseudo-spectrum; (b) Positioning pseudo-spectrum computed with toroidal pulses at a signal-to-noise ratio of 20 dB; (c) Experimental positioning pseudo-spectrum constructed from measured data. The target is located at $(x, y, z) = (0, 0.1, 0.2)$ m, with the toroidal pulses propagating in the $z$-direction and their cross-sectional center located at $(x = 0, y = 0)$.

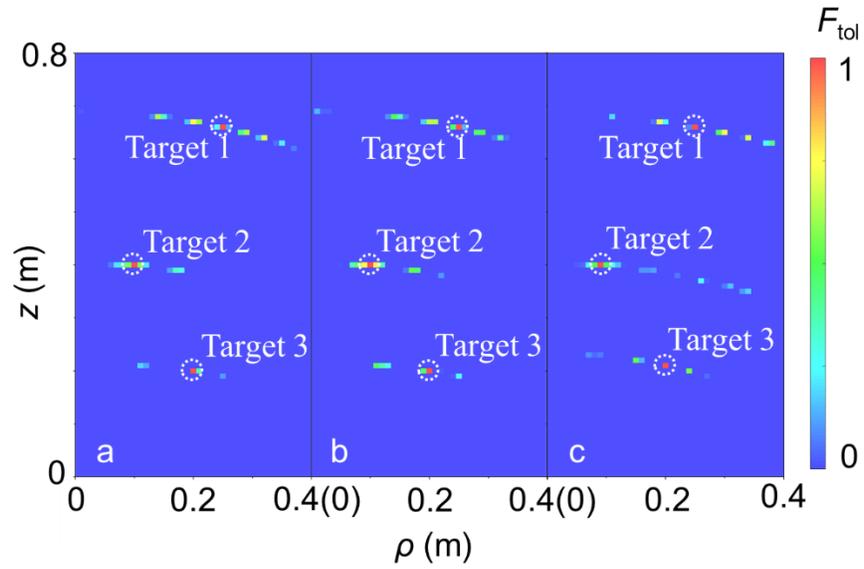

**Fig. 3 Multi-target positioning pseudo-spectrum based on toroidal pulses.** (a) Positioning pseudo-spectrum computed using ideal toroidal pulses; (b) Positioning pseudo-spectrum computed using toroidal pulses with a signal-to-noise ratio of 20 dB; (c) Positioning pseudo-spectrum constructed from experimental data.

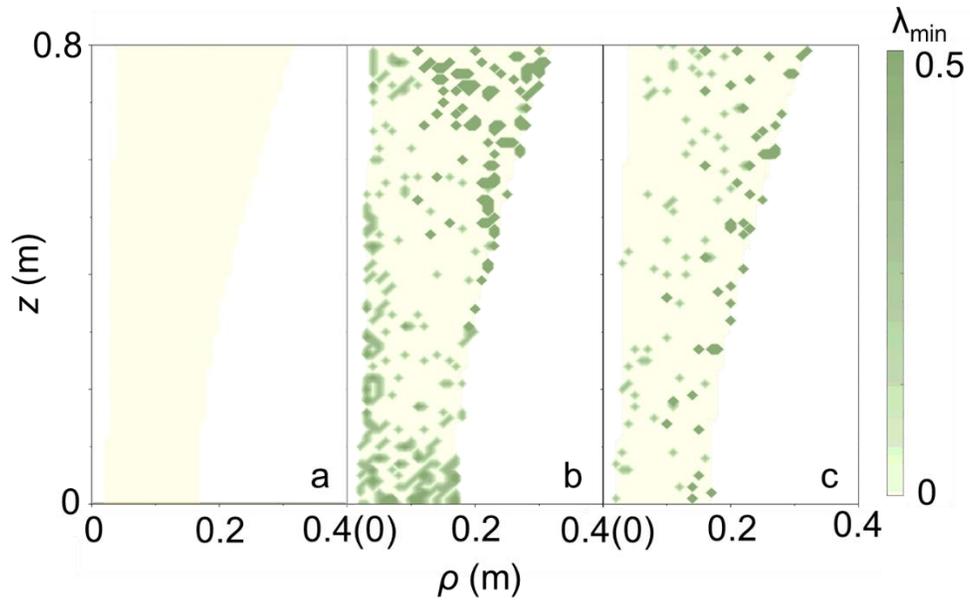

**Fig. 4 Positioning displacement metrics.** (a) Displacement metrics computed using ideal toroidal pulses; (b) Displacement metrics computed using toroidal pulses with a signal-to-noise ratio of 20 dB; (c) Displacement metrics constructed from experimental data.

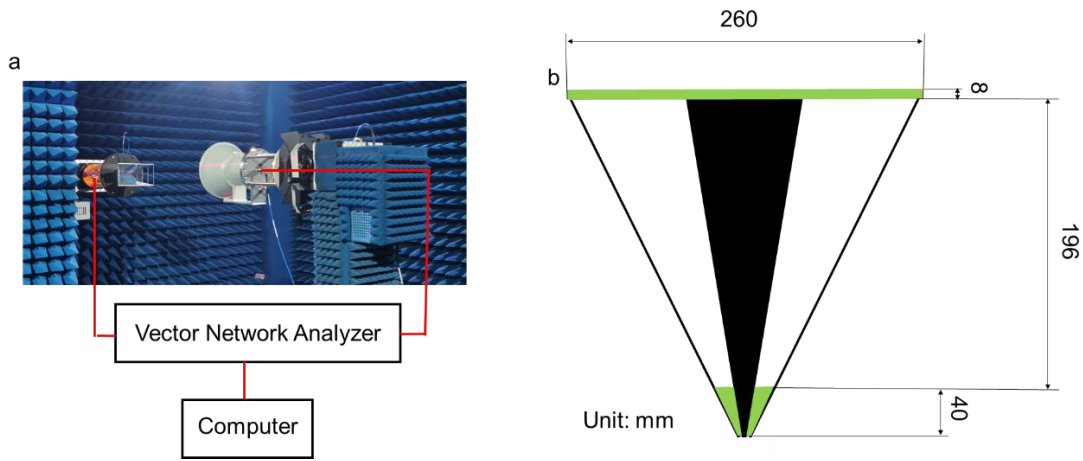

**Fig. 5 (a) Experimental setup and (b) schematic of coaxial horn antenna transmitting microwave toroidal pulses.** The experimental setup utilized coaxial horn antennas as the transmitting antennas for toroidal fields, while double-ridged waveguide horn antennas served as the dual-polarized receiving antennas, simulating the user positionings to be localized.